\documentclass[twocolumn,preprintnumbers,superscriptaddress,floatfix]{revtex4}
\usepackage{amsmath}
\usepackage{amssymb}
\usepackage{mathrsfs}
\usepackage{graphicx}
\usepackage{bm}
\usepackage{epstopdf}
\usepackage[colorlinks=true,citecolor=blue,linkcolor=blue,urlcolor=blue]{hyperref}
\usepackage{bookmark}
\usepackage{color}
\usepackage{subfigure}
\usepackage{graphicx}

\usepackage{xcolor,color}

\begin{document}
\title{Nonlinear magneto-optical response across van Hove singularity in  a non-centrosymmetric magnetic Weyl semimetal}


\author{Jian Li }
\affiliation{School of Physics and Electronic Science, Changsha University of Science and Technology, Changsha 410076, China}
\affiliation{Provincial Key Laboratory of Informational Service for Rural Area of Southwestern Hunan, Shaoyang University, Shaoyang 422000, China}
\affiliation{Hunan Province Higher Education Key Laboratory of Modeling and Monitoring on the Near-Earth Electromagnetic Environments (Changsha University of Science and Technology)}

\author{Kai-He Ding}\email{dingkh@csust.edu.cn}
\affiliation{School of Physics and Electronic Science, Changsha University of Science and Technology, Changsha 410076, China}

\author{Lijun Tang}\email{tanglj@csust.edu.cn}
\affiliation{School of Physics and Electronic Science, Changsha University of Science and Technology, Changsha 410076, China}
\begin{abstract}
We investigate the nonlinear magneto-optical response in non-centrosymmetric magnetic Weyl semimetals featuring a quadratic tilt, focusing particularly on the influence of the van Hove singularity (VHS). In the absence of a magnetic field,
the second-order nonlinear Drude conductivity components exhibit inflection or dip behavior across the VHS. In contrast, the second-order nonlinear anomalous Hall conductivity, primarily governed by the Berry curvature dipole, manifests a subtle plateau-like structure. As the tilt strength increases, the VHS energy escalates, thereby amplifying the VHS-induced characteristics within these second-order conductivity components.
However, in the presence of a magnetic field, we show that the resultant magnetic moment suppresses nonlinear electron transport while enhancing
nonlinear hole transport. 
This effect serves to mitigate the impact of the VHS, resulting specifically in an asymmetric peak or a kinked-like structure in the magnetic field-induced contribution to the second-order nonlinear conductivity near the Weyl nodes. These findings provide new insights into the intricate interplay among the VHS, Berry curvature, and magnetic moment in nonlinear magneto-optical transport through non-centrosymmetric magnetic Weyl semimetals.

\end{abstract}

\maketitle

\section{Introduction}
The field of condensed matter physics has witnessed a significant surge in interest and investigation into a unique class of materials known as Weyl semimetals (WSMs)\cite{Burkov2011,Shin-Ming2015}. These materials exhibit a specific electronic structure characterized by the presence of  band-crossing points, termed Weyl nodes\cite{lvsci2015,Bucciantini2017}. These nodes introduce a new kind of quasiparticle, Weyl fermions, distinguished by their linear dispersion relations and fundamental properties, making them pivotal entities in the realm of topological materials \cite{Weng15,Belopolski15}.

One of the striking features of WSMs lies in the Berry curvature flux emanating from these Weyl nodes within the Brillouin zone, behaving like magnetic fields in momentum space.
These nodes serve as sources and sinks of Berry curvature flux, intimately tied to the chiralities of the Weyl fermions \cite{Bentmann2021,Xiao10}. The distinct topological electronic structure of WSMs has paved the way for the exploration of unconventional phenomena and emergent behaviors, such as high mobility effects \cite{timuskprb2013,orlitanp2014}, Fermi arcs \cite{Burkov2011,xuprl2011,lvsci2015,wanprb2011,Belopolski15}, and the intriguing chiral anomaly \cite{Xiong15, Shiva20, Kipp21,Yuan20, Xiaochun15}. These phenomena have spurred considerable interest and investigation into the transport properties of these materials \cite{Tabert2016,Armitage18,Jin-Feng2022}.
The simultaneous application of electric and magnetic fields to WSMs has unveiled underlying transport mechanisms,
yielding notable discoveries like positive longitudinal magnetoconductivity\cite{yangnp2015,shekharnp2015,Xiong15,linc2015,wangnc2016,lvprl2017,linc2016,dengprl2013} and giant planar Hall effect \cite{burkovprb2017,nandyprl2017}. 
Further research has extended into exploring the nonlinear magneto-optical responses to external electric fields, revealing intriguing phenomena such as the quantum nonlinear Hall effect, which arises solely from the dipole moment of the Berry curvature in the absence of an applied magnetic field \cite{Morimoto16,Hai-Zhou18,FuruZhang20, Sodemann15}.

The foundational understanding of WSMs traces back to the violation of symmetries. According to the Nielsen-Ninomiya theorem \cite{Nielsen81,Nielsen83}, the existence of WSMs necessitates the breaking of time-reversal or inversion symmetry\cite{Zyuzin12}. Investigations into the impact of breaking these symmetries have unveiled transitions from Dirac semimetals to WSMs, manifesting in various transport signatures and anomalous effects \cite{Zhijun12,Zyuzin12,Armitage18}. The manipulation of these symmetries through external means, such as introducing the magnetic atoms or the magnetically doped multilayer heterostructure, 
has led to the realization and study of magnetic WSMs\cite{liusci2019,puphalprl2020, yinnat2018,LiuE.,morsci2019,Guoqing18,Cong Li2023}, showing fascinating effects like exotic drumhead surface states\cite{yinnat2018}, chiral magnetic effects and the giant anomalous Hall effect\cite{LiuE.,morsci2019}.

In certain WSMs, such as the TaAs family, the convergence of Weyl nodes gives rise to a van Hove singularity (VHS) at relatively low energies \cite{YONG HU2022,Zhenyu2022,Ebad-Allah2023}. While the impact of VHS on conventional linear transport, including the emergence of negative magnetic resistance induced by the VHS\cite{schumannprb2017,dingprb2023}, has been elucidated, its implications in nonlinear magneto-optical responses have been largely overlooked.

In this work, we investigate the nonlinear magneto-optical response across the VHS within non-centrosymmetric magnetic Weyl semimetals. We find that depending on the coupling between the conventional velocity and Berry curvature dipole, the VHS induces inflection points, dip behaviors, and subtle plateau-like structures in the second-order nonlinear conductivity components at zero magnetic field. Furthermore, we demonstrate that the application of a magnetic field suppresses nonlinear electron transport while enhancing nonlinear hole transport due to the presence of the magnetic moment. This duality mitigates the influence of the VHS but also results in the emergence of asymmetric peaks or kink-like structures in the magnetic field-induced contribution to the second-order nonlinear conductivities near the Weyl nodes.
Upon evaluating the magnitude of these nonlinear conductivity components, our findings indicate the potential observability of VHS-related features in non-centrosymmetric magnetic WSMs subjected to simultaneous electric and magnetic fields.


This paper is organized as follows: In Sec. II, we introduce a theoretical model describing non-centrosymmetric magnetic WSMs in the presence of the VHS.
In Sec. III, we establish the magneto-optical transport equations within the semiclassical approximation, considering the simultaneous impact of electric and magnetic fields.
In Sec. IV, we present analytical formulations for second-order nonlinear magneto-optical conductivities, followed by a detailed analysis of their numerical results.
Finally, we conclude in Sec. V.

\section{Theoretical model}
\label{sec:Hamiltonian Models for Weyl semimetals}

A non-centrosymmetric magnetic WSM can be effectively characterized through a low-energy Hamiltonian \cite{Rui-Hao21,Hai-Zhou10}:
\begin{equation}
H = v_F[k_x\sigma_x s_z - k_y\sigma_y - \lambda (k_z^2 - Q_D^2)\sigma_z ] + H_{B},
\label{Ham}
\end{equation}
where $\sigma_i$ and $s_i$ ($i=x,y,z$) represent the Pauli matrices acting on the orbital and spin space, $k_i$ denotes the wave vector, and $v_F$ is the Fermi velocity. The term $\lambda (k_z^2 - Q_D^2)\sigma_z$ introduces two Dirac nodes located at $\mathbf{k}=(0,0,\pm Q_D)$. The Hamiltonian $H_B$ consists of three components:
\begin{equation}
H_B = Rk_z^2+V_I k_z\sigma_z s_z+ J_{ex} s_z,
\label{Ham2}
\end{equation}
where $V_I k_z \sigma_z s_z$ denotes the inversion symmetry-breaking term, leading to the splitting of each Dirac node into two Weyl nodes with opposite chirality along the $z$-axis. The term $Rk_z^2$ disrupts the particle-hole symmetry(PHS), resulting in a tilted energy dispersion around the Weyl nodes and modifying their energy.
It is important to emphasize that in Weyl semimetals hosting Fermi pockets, particularly when these pockets are not directly linked to the Weyl nodes, the inclusion of a PHS-breaking term is crucial for accurately describing this phenomenon\cite{Rui-Hao21,nagprb2022,msprb2020}. The formulation of this PHS-breaking term can be achieved by adjusting the lattice structure (see, for example, Refs.\cite{Zhijun12, dipprb2020}). In the presence of the D4h symmetry of the crystal\cite{yqprb2020}, the PHS-breaking term is characterized by its proportionality to $k_z^2$ at the lowest order momentum\cite{Rui-Hao21}. 
However, at low energy, higher-order terms such as cubic or quartic terms have minimal influence and can be disregarded. Furthermore, by expanding the energy dispersion around the Weyl nodes and retaining terms up to the first order in $k_z$, one can derive the energy dispersion in the vicinity of the Weyl node of chirality. This representation corresponds to the low-energy excitations observed in materials such as MoTe$_2$ and WTe$_2$, featuring a linear term proportional to $k_z$, indicating a tilt in the Weyl cone\cite{dasprb2022}.
Finally, $J_{ex} s_z$ represents the time-reversal symmetry-breaking term, arising from the exchange interaction between the Weyl-fermion spin and the magnetization in a magnetic WSM.

\begin{figure}[htbp]
\centering
\includegraphics[width=8.2cm]{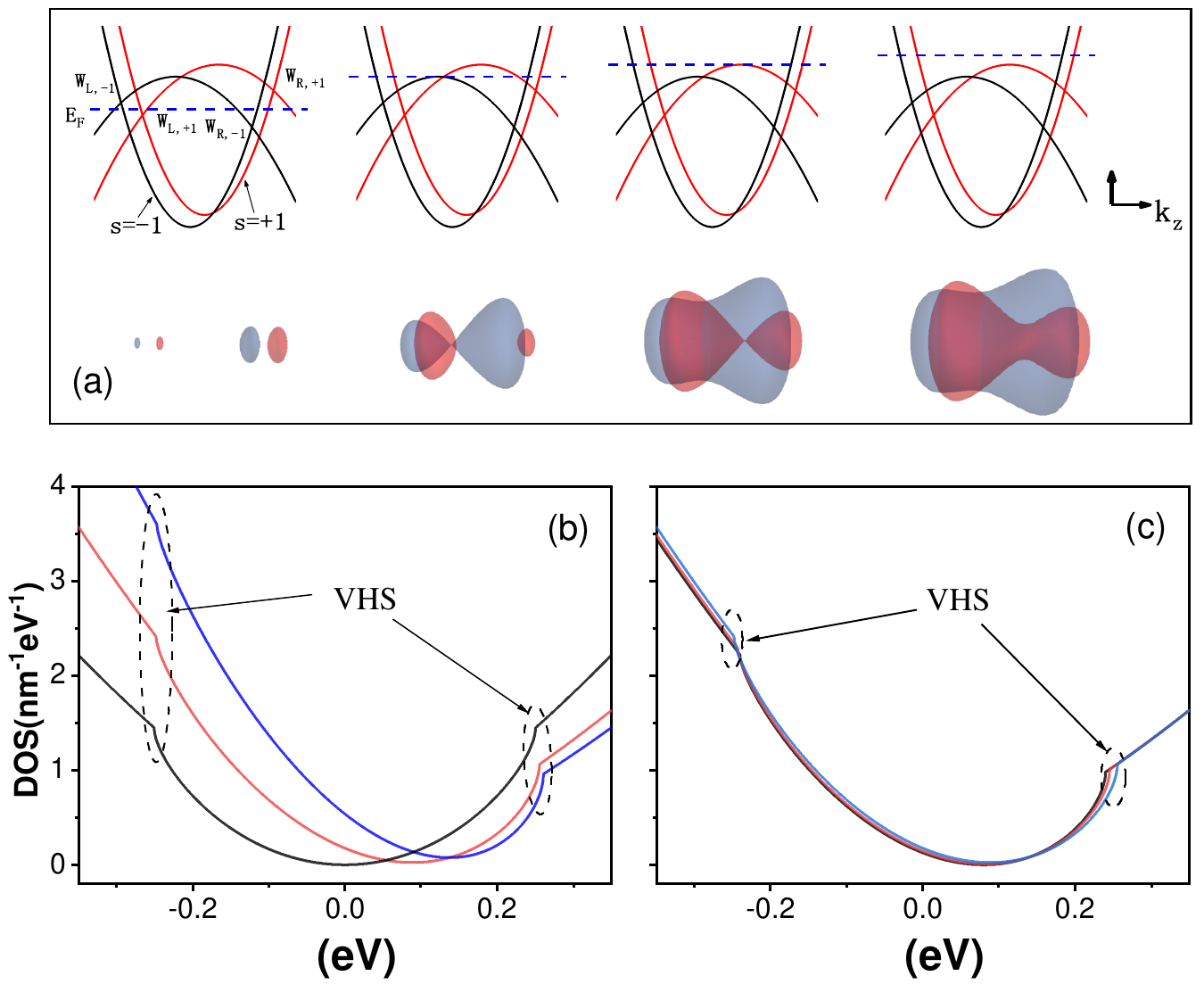}
\caption{(color online) (a) The schematic portrays energy band dispersion with four curves forming two pairs labeled by $s=\pm1$. Each pair intersects, creating four Weyl nodes denoted as $W_{\Lambda,s}(\Lambda=L,R)$. The Fermi energy $E_F$, indicated by the dashed blue line, traverses the band curves, progressively increasing from the left to the right-hand side of the upper panel. Lower panel of figure (a): Representation of the corresponding Fermi surface, exhibiting asymmetric features and undergoing a Lifshitz transition\cite{lifjetp1960} as the Fermi energy increases. The density of states for the tilt $R=0eV\cdot nm^2$(black line), $R=0.02eV\cdot nm^2$(red line), and $R=0.04eV\cdot nm^2$(blue line) at $V_I=0.05eV\cdot nm$ (b), and for the parameter $V_I=0eV\cdot nm$(black line), $V_I=0.03eV\cdot nm$(red line), and $V_I=0.05eV\cdot nm$(blue line) at $R=0.02eV\cdot nm^2$ (c). The other parameters are set as $Q_D=2nm^{-1}$, $J_{ex}=0eV$, and $\lambda=0.06nm$. }
\label{Fig1}
\end{figure}
Diagonalizing Hamiltonian (\ref{Ham}) yields the energy spectrum:
\begin{equation}
\varepsilon_{\alpha}^{s} = \alpha v_F \sqrt{k_\parallel^2 + \Delta_s^2(k_z)}  +Rk_z^2+ s J_{ex},
\label{energy3}
\end{equation}
where $k_\parallel^2 = k_x^2 + k_y^2$, $\Delta_s(k_z) = \lambda (k_z^2 - Q_D^2) - s V_I k_z/v_F$, and $\alpha=\pm$ denote the conduction or valence bands. The corresponding eigenvectors are:
\begin{equation}
|u_{\mathbf{k}+}^s\rangle = \left(\begin{array}{cc} \cos\frac{\theta_s}{2} & e^{i\phi_s} \sin\frac{\theta_s}{2}
\end{array}\right)^t,
\label{wav1}
\end{equation}
\begin{equation}
|u_{\mathbf{k}-}^s\rangle = \left(\begin{array}{cc} e^{i\phi_s} \sin\frac{\theta_s}{2} & -\cos\frac{\theta_s}{2}
\end{array}\right)^t,
\label{wav2}
\end{equation}
where $\cos\theta_s = \frac{\Delta_s(k_z)}{\sqrt{k_\parallel^2  + [\Delta_s(k_z)]^2}}$, and $\tan\phi_s = -\frac{sk_y}{ k_x}$.

In Eq. (\ref{energy3}), $s=\pm1$ denotes the chirality of the Weyl node. This can be identified through the Berry curvature $\mathbf{\Omega}_{\alpha}^{s}$, expressed as:
\begin{equation}
\mathbf{\Omega}_{\alpha}^{s} = -\operatorname{Im}\left[ \langle\nabla_\mathbf{k} u_{\mathbf{k}\alpha}^s| \times |\nabla_\mathbf{k} u_{\mathbf{k}\alpha}^s\rangle \right].
\label{berv}
\end{equation}
By substituting Eqs. (\ref{wav1}) and (\ref{wav2}) into Eq. (\ref{berv}), the resulting equations are:
\begin{equation}
\Omega_{\alpha x/y}^{s} = - \frac{s\alpha k_{x/y} \Delta_s'(k_z)}{{2K^3}}, \
\Omega_{\alpha z}^{s} = - \frac{{s\alpha\Delta_s(k_z)}}{{2K^3}},
\label{berrv}
 \end{equation}
where $\Delta_s'(k_z)=\partial \Delta_s(k_z)/\partial k_z$, and $K=(k_x^2+k_y^2+\Delta_s^2)^{1/2}$.

The dispersion relation (\ref{energy3}) depicted in Fig. \ref{Fig1}(a) encompasses two pairs of band curves, denoted by $s=\pm1$. Each band pair intersects, forming two Weyl nodes located at $(0,0,\pm\sqrt{Q_D^2 + q^2} - s q)$ with $q=V_I/(2\lambda v_F)$, labeled as $W_{\Lambda,s}$ ($\Lambda=L,R$) (see upper panel of Fig. \ref{Fig1}(a)).
When the Fermi energy crosses the bands at energy $\varepsilon>0$, an increase leads to a Lifshitz transition\cite{lifjetp1960}, transforming the Fermi surface from two disconnected parts into a single one(see lower panel of Fig.\ref{Fig1}(a)). Throughout this transition, the Fermi surface consistently remains asymmetric. These distinct Fermi surface behaviors, coupled with the chiral anomaly, significantly impact the nonlinear magneto-optical responses in WSMs (as discussed below).
Fig.\ref{Fig1}(b) illustrates the density of states (DOS) for varying tilt $R$. A detailed calculation of DOS is provided in the appendix. Evidently, the DOS displays inflections at VHS, which remain symmetrical with respect to the zero-energy point at tilt $R=0$. However, with the application of tilt, the symmetry of DOS is lost due to the PHS breaking. The asymmetry induced by $R$ can be further heightened by the breaking of inversion symmetry (see Fig. \ref{Fig1}(c)).

\section{Magneto-optical transport equations within Semiclassical approximation }
\label{motion}
We consider the simultaneous application of the static magnetic field ${\bf{B}}$ and light field ${\bf{E}}$ to the system. The resultant electric current can be computed using the following integral:
\begin{align}
{\bf{j}} =  - e\int {[d{\bf{k}}]} D{\bf{\dot r}}f({\bf{k}},{\bf{r}},t),\label{cur}
\end{align}
where ${\rm{[d}}{\bf{k}}] = {{d{\bf{k}}} / {{{(2\pi )}^3}}}$, and the weighting factor $D$ emerges due to the alteration of the phase volume caused by the electric and magnetic fields (see below). ${\bf{\dot r}}$ represents the electron's velocity and is obtained by solving the following equation of motion \cite{Xiao10}:
\begin{equation}
\begin{aligned}
&{\bf{\dot r}} = {1 \over \hbar }{\nabla _{\bf{k}}}{\epsilon _{\alpha\bf{k}}^{s}} - {\bf{\dot k}} \times {{\bf{\Omega }}_{\alpha}^{s}}, \\
&\hbar {\bf{\dot k}} =  - e{\bf{E}} - e{\bf{\dot r}} \times {\bf{B}}.\label{mot}
\end{aligned}
\end{equation}
This equation describes the trajectory of an electron in phase space, viewed as a wave packet within the semiclassical approximation. The self-rotation of the wave packet around its center of mass under the magnetic field induces the existence of an orbital magnetic moment ${{\bf{m}}_{\alpha}^s}$, leading to a modification of the dispersion relation (\ref{energy3}):
\begin{equation}
\epsilon _{\alpha\bf{k}}^{s} = \varepsilon_\alpha^s - {{\bf{m}}_{\alpha}^{s}} \cdot {\bf{B}}\label{mkds}
\end{equation}
 with
\begin{equation}\label{mk}
{{\bf{m}}_{\alpha}^{s}} =  - {e \over {2\hbar }}{\mathop{\rm Im}\nolimits} [\left\langle {{\nabla _{\bf{k}}}{u_{\bf{k}\alpha}^s}} \right| \times (H - \varepsilon _{\alpha}^s)\left| {{\nabla _{\bf{k}}}{u_{\bf{k}\alpha}^s}} \right\rangle ].
\end{equation}
By substituting the wave function $|u_{\mathbf{k}\alpha}^s\rangle$ in Eq. (\ref{mk}), the orbit magnetic moment are further expressed as
\begin{equation}
\begin{aligned}
m_{\alpha x/y}^{s} =  - s\alpha e{v_F}{{{k_{x/y}}\Delta_s '({k_z})} \over {2{K^2}}},
m_{\alpha z}^{s} =  - s\alpha e{v_F}{{\Delta_s ({k_z})} \over {2{K^2}}}.
\end{aligned}
\end{equation}
 By solving the coupled equations (\ref{mot}),
one can obtain:
\begin{equation}
{D\bf{\dot r}} = \mathbf{v}_{\alpha}^s + \frac{e}{\hbar}{\bf{E}} \times {{\bf{\Omega }}_{\alpha}^{s}} + {e \over \hbar }(\mathbf{v}_{\alpha}^s \cdot {{\bf{\Omega }}_{\alpha}^{s}}){\bf{B}},\label{mot1}
\end{equation}
\begin{equation}
{D\bf{\dot k}} =-\frac{ e}{\hbar}{\bf{E}} - {e \over \hbar }\mathbf{v}_{\alpha}^s \times {\bf{B}} - {{{e^2}} \over \hbar^2 }({\bf{E}} \cdot {\bf{B}}){{\bf{\Omega }}_{\alpha}^{s}}, \label{mot2}
\end{equation}
where $\mathbf{v}_{\alpha}^s=\frac{1}{\hbar}\nabla _{\bf{k}}\epsilon _{\alpha\bf{k}}^{s}$, and
$
D = 1 + {e \over \hbar }{\bf{B}} \cdot {{\bf{\Omega }}_{\alpha}^{s}}.
$
In Eq.\eqref{mot1}, the first term is the usual group velocity, 
the second term is  anomalous velocity induced by the Berry curvature, and the third term represents their coupling induced by the magnetic field. In Eq.\eqref{mot2}, the first two terms are the usual Lorentz force, while the last term is associated with chiral anomaly \cite{Nielsen83}.

In Eq.\eqref{cur}, $f({\bf{k}},{\bf{r}},t)$ is a distribution function obeying semi-classical Boltzmann kinetic equation as follows:
\begin{align}\label{frkt}
{{{\rm{d}}f({\bf{k}},{\bf{r}},t)} \over {dt}} = {{\partial f} \over {\partial t}} + {\bf{\dot k}}{{\partial f} \over {\partial {\bf{k}}}} + {\bf{\dot r}}{{\partial f} \over {\partial {\bf{r}}}} = {I_c}\{ f\}
\end{align}
with ${I_c}\{ f\}$ a collision integral term.
We assume that the light field has the form of ${\bf{E}}(t) = {\bf{E}}{e^{ - i\omega t}}$, and the magnetic field is homogeneously applied on WSM. In this situation, the distribution function $f({\bf{k}},{\bf{r}},t)$ becomes independent of the spatial coordinate. Within the relaxation time approximation, Eq.\eqref{frkt} reduces to
\begin{align}\label{Boltzmann}
{{\partial f} \over {\partial t}} + {\bf{\dot k}}{{\partial f} \over {\partial {\bf{k}}}} = {\rm{ - }}{{f - {f_0}} \over \tau },
\end{align}
where $f_0$ is the Fermi distribution function. For simplicity, we ignore internode scattering, thus considering $\tau$ from Eq.\eqref{Boltzmann} as the  intranode scattering time. Detailed discussions on the influence of internode scattering in the nonlinear transport regime are available in Ref.\cite{dasprb2022}.
To solve Eq.(\ref{Boltzmann}), the distribution function $f$ is expanded up to second order in the electric field:
\begin{align}\label{component}
f = {f_0} + {f_1}{e^{ - i\omega t}} + {f_2}{e^{ - 2i\omega t}}.
\end{align}
 Inserting Eq. \eqref{component} into Eq.\eqref{Boltzmann} and equating equal powers of $\mathbf{E}$, we get the recursion equations.
Further solving these equations finally leads to
\begin{align}\label{f1}
{f_1} =  {\tau  \over {1 - i\omega \tau }}{1 \over {\hbar D}}[  e{\bf{E}} + {{{e^2}} \over \hbar }({\bf{E}} \cdot {\bf{B}}){{\bf{\Omega }}_{\alpha}^{s}}] \cdot {\nabla _{\bf{k}}}{f_0},
\end{align}
\begin{footnotesize}
\begin{equation}
\begin{aligned}
&{f_2} = {{{\tau ^2}} \over {(1 - i\omega \tau )(1 - 2i\omega \tau )}}{({1 \over {\hbar D}})^2} \\
&\times[  e{\bf{E}} + {{{e^2}} \over \hbar }({\bf{E}} \cdot {\bf{B}}){{\bf{\Omega }}_{\alpha}^{s}}] \cdot {\nabla _{\bf{k}}}\{ [  e{\bf{E}} + {{{e^2}} \over \hbar }({\bf{E}} \cdot {\bf{B}}){{\bf{\Omega }}_{\alpha}^{s}}] \cdot {\nabla _{\bf{k}}}{f_0}\}.
\end{aligned}
\label{f2}
\end{equation}
\end{footnotesize}
Substituting Eqs.(\ref{f1}) and (\ref{f2}) in Eq.(\ref{cur}), we will get the analytic expressions of the electric current in powers of $\mathbf{E}$(see below), which provides a basis for exploring the nonlinear magneto-optical transport through WSM.

\section{Second-order nonlinear magneto-optical conductivities}
\label{sec:General expression for SHG}
We now explore the second-order
nonlinear magneto-optical response of WSMs.
Employing Eq.(\ref{cur}),
the expression for the second-order nonlinear current response at the frequency $2\omega$ is given by
\begin{equation}
\mathbf{j}^s =  - e\int {[d{\bf{k}}]} D{\bf{\dot r}}({f_1} + {f_2}).\label{currt}
\end{equation}
By substituting Eq.(\ref{mot1}) into Eq.(\ref{currt}), the second-order nonlinear current can be expressed as:
\begin{align}\label{desti}
\mathbf{j}^s =  - e\int {[d{\bf{k}}]\{ [{{\bf{v}}_{\alpha}^s} + {e \over \hbar }({{\bf{v}}_{\alpha}^s} \cdot {{\bf{\Omega }}_{\alpha}^s}){\bf{B}}]{f_2} + {e \over \hbar }{\bf{E}} \times {{\bf{\Omega }}_{\alpha}^s}{f_1}\} }.
\end{align}
Combining Eqs.(\ref{f1}), (\ref{f2}) with Eq.(\ref{desti}) yields the following expressions:
\begin{widetext}
\begin{equation}
\begin{aligned}
&\mathbf{j}^s =  - {{e\tau } \over {(1 - 2i\omega \tau )}}\int  { [d{\bf{k}}]\over {\hbar D}}
[{{\bf{v}}_{\alpha}^s} + {e \over \hbar }({{\bf{v}}_{\alpha}^s} \cdot {{\bf{\Omega }}_{\alpha}^{s}}){\bf{B}}]
[e{\bf E} + {{{e^2}} \over \hbar }({\bf{E}} \cdot {\bf{B}}){\bf{\Omega }_{\alpha}^{s}}] \cdot {{\partial {f_1}} \over {\partial {\bf{k}}}}\\
&-{{{e^2}\tau } \over {\hbar (1 - i\omega \tau )}}\int {[d{\bf{k}}] \over {\hbar D}} {\bf{E}} \times {{\bf{\Omega }}_{\alpha}^{s}}[e{\bf{{\bf E}}} + {{{e^2}} \over \hbar }({\bf{E}} \cdot {\bf{B}}){{\bf{\Omega }}_{\alpha}^{s}}] \cdot {{\partial {f_0}} \over {\partial {\bf{k}}}}.
\end{aligned}
\label{cdestit}
\end{equation}
\end{widetext}
From Eq. (\ref{cdestit}), it is apparent that, alongside the $\mathbf{E} \times \mathbf{\Omega}_{\alpha}^s$ term, the chiral anomaly (i.e., the $\mathbf{E} \cdot \mathbf{B}$ term) contributes to the Hall current. Moreover, the existence of a magnetic field induces a magnetic moment (refer to Eq. \eqref{mk}), exerting distinct effects on the transport properties in the electron and hole regions(detailed below).

\subsection{Second-order nonlinear conductivity without magnetic field}

 In the absence of a magnetic field($B=0$), Eq.\eqref{cdestit} simplifies as follows:
\begin{equation}
\begin{array}{cll}
\mathbf{j}^s &=& -\kappa_1\int [d\mathbf{k}] \mathbf{v}_{\alpha}^se\mathbf{E} \cdot \frac{\partial}{ \hbar\partial\mathbf{k}}[(e\mathbf{E} \cdot \mathbf{v}_{\alpha}^s)\frac{\partial f_0^s}{\partial\varepsilon_{\alpha}^s}]\\
&&- \kappa_2\int [d\mathbf{k}] \mathbf{E} \times \mathbf{\Omega}_{\alpha}^s(e\mathbf{E} \cdot \mathbf{v}_{\alpha}^s)\frac{\partial f_0^s}{\partial\varepsilon_{\alpha}^s},
\end{array} \label{j0b1-modified}
\end{equation}
where ${\kappa _1} = {{e{\tau ^2}} \over {(1 - 2i\omega \tau )(1 - i\omega \tau )}}$ and ${\kappa _2} = {{{e^2}\tau } \over {\hbar (1 - i\omega \tau )}}$. We express Eq.\eqref{j0b1-modified} in the form of $j_a^s=\sigma_{abc}^sE_b(\omega)E_c(\omega)$, where $\sigma_{abc}^s$ is the second-order nonlinear conductivity:
\begin{equation}
\sigma_{abc}^s(2\omega) = \sigma_{abc}^{s,0} + \sigma_{abc}^{s,H}.
\label{curzerob-modified}
\end{equation}
In Eq.\eqref{curzerob-modified}, the first term denotes the nonlinear Drude conductivity: 
\begin{equation}
\sigma_{abc}^{s,0} = \frac{e^2\kappa_1}{\hbar}\int [d\mathbf{k}]\frac{\partial v_{\alpha a}^s}{\partial k_b} v_{\alpha c}^s \frac{\partial f_0^s}{\partial\varepsilon_{\alpha}^s},
\label{j0b-modified}
\end{equation}
while the second term is the nonlinear anomalous Hall conductivities induced by the intrinsic
Berry curvature dipoles: 
\begin{equation}
{\sigma^{s,H}_{abc}} = -\varepsilon_{adc}e\kappa_2\int [d\mathbf{k}]\Omega^s_{\alpha d} v^s_{\alpha b} \frac{\partial f_0^s}{\partial\varepsilon_{\alpha}^s}, \label{hde1-modified}
\end{equation}
where $\varepsilon_{adc}$ represents the three-dimensional Levi-Civita antisymmetric tensor, and the integral denotes the Berry curvature dipoles\cite{Sodemann15}.

 \subsubsection{Second-order nonlinear conductivity $\sigma_{abc}^{s,0}$}
Since the velocities $\mathbf{v}_{\alpha}^s$ exhibit odd functional dependencies on the momentum components $\mathbf{k}$, the integrands in Eq.\eqref{j0b-modified}$ $ maintain consistent odd symmetries regarding the momentum $k_x$, $k_y$, or $k_z$.
In the absence of a tilt in the Weyl cone, the Fermi energy surface preserves its symmetry relative to the origin. In this context, contributions to the electric current from both positive($W_{R,s}$) and negative($W_{L,s}$) Weyl nodes possess equal magnitudes but opposite directions, resulting in the complete nullification of the conductivity $\sigma_{abc}^{s,0}$.
However, when considering the tilt term $Rk_z^2$, the parity characteristics of the velocity $\mathbf{v}_\mathbf{k}^s$ remain unchanged. Nonetheless, the Fermi surface symmetry is exclusively disrupted along the $z$ direction while remaining preserved in the $x$ and $y$ directions.
Thus, non-zero components of the conductivity tensor $\sigma_{abc}^{s,0}$ necessitate an even number of $x$ or $y$ indices, as outlined below:
\begin{footnotesize}
\begin{equation}
\begin{aligned}
\sigma _{zxx}^{s,0} =\frac{e^2v_F\kappa_1}{8\pi^2\hbar^2 }\int {d{k_z}} \frac{\Delta_s (k_z)\Delta_s '(k_z)}{r (k_z)}[1 - \frac{\Delta_s^2(k_z)}{r^2 (k_z)}],
\end{aligned}\label{szzz1}
\end{equation}
\end{footnotesize}
\begin{footnotesize}
\begin{equation}
\begin{aligned}
&\sigma _{xzx}^{s,0} =   {{{e^2}{v_F}{\kappa _1}} \over {8{\pi ^2}{\hbar ^2}}}\int {d{k_z}} \{ [ \frac{\Delta_s (k_z)\Delta_s '(k_z)}{r (k_z)} + {{2 R{k_z}} \over {\hbar {v_F}}}]\\
&\times [-1 - {{\Delta _s^2({k_z})} \over {{r^2}({k_z})}}]\},
\end{aligned}
\end{equation}
\end{footnotesize}
\begin{footnotesize}
\begin{equation}
\begin{aligned}
&\sigma _{zzz}^{s,0} =  -\frac{e^2v_F\kappa_1}{4\pi^2\hbar^2 }\int {d{k_z}} \{[\frac{\Delta_s (k_z)\Delta_s '(k_z)}{r (k_z)}+\frac{2R k_z}{\hbar v_F}]\\
&\times [ 2\lambda \Delta_s(k_z) +[\Delta_s '(k_z)]^2 - \frac{\Delta_s^2 ({k_z})[\Delta_s '(k_z)]^2}{r^2 (k_z)}] \\
& + {{2R} \over \hbar v_F}[\Delta_s ({k_z})\Delta_s '({k_z}) +r(k_z) \frac{2Rk_z}{\hbar v_F}]\},
\end{aligned}\label{szzz2}
\end{equation}
\end{footnotesize}
where $r ({k_z}) = {{\mu  - s{J_{ex}} - {\rm{R}}k_z^2} \over {\hbar {v_F}}}$. The Fermi surface symmetry concerning the $k_x$ and $k_y$ axes implies that the other components satisfy the relation $\sigma _{xxz}^{s,0} = \sigma _{zyy}^{s,0} = \sigma _{yyz}^{s,0} = \sigma _{zxx}^{s,0}$, and $\sigma _{yzy}^{s,0}=\sigma _{xzx}^{s,0} $.
It is noted that $\sigma_{xxz}$ does not precisely align with the component $\sigma_{xzx}$, while it exhibits an approximate equivalence to $\sigma_{xzx}$ particularly under conditions of small tilt amplitudes, denoted by $R$.
Additionally,
Eqs. \eqref{szzz1} and \eqref{szzz2} can be expressed as follows:
$\sigma _{zxx}^{s,0} =\frac{e^3v_F\sigma_{d1}}{8\pi^2\hbar^2 }{{{\tau ^2}} \over {(1 - 2i\omega \tau )(1 - i\omega \tau )}}$
and
$\sigma _{zzz}^{s,0} =-\frac{e^3v_F\sigma_{d2}}{4\pi^2\hbar^2 }{{{\tau ^2}} \over {(1 - 2i\omega \tau )(1 - i\omega \tau )}}$.
Here, $\sigma_{d1}$ and $\sigma_{d2}$ represent the integral terms in Eqs. \eqref{szzz1} and \eqref{szzz2}, respectively. It is evident that these parameters, $\sigma_{d1}$ and $\sigma_{d2}$ are not dependent on the frequency $\omega$.
In the transport limit, where $\omega\tau << 1$, the conductivity components $\sigma_{zxx}^{s,0}$ and $\sigma_{zzz}^{s,0}$ are proportional to $\tau^2$ and independent of frequency. Conversely, in the optical or clean limit, where $\omega\tau >> 1$, these conductivity components are proportional to $1/\omega^2$ and independent of $\tau$. These results are consistent with those of the single-node model\cite{Gaoyang22}.

\begin{figure}[htbp]
\centering
\includegraphics[width=7.6cm]{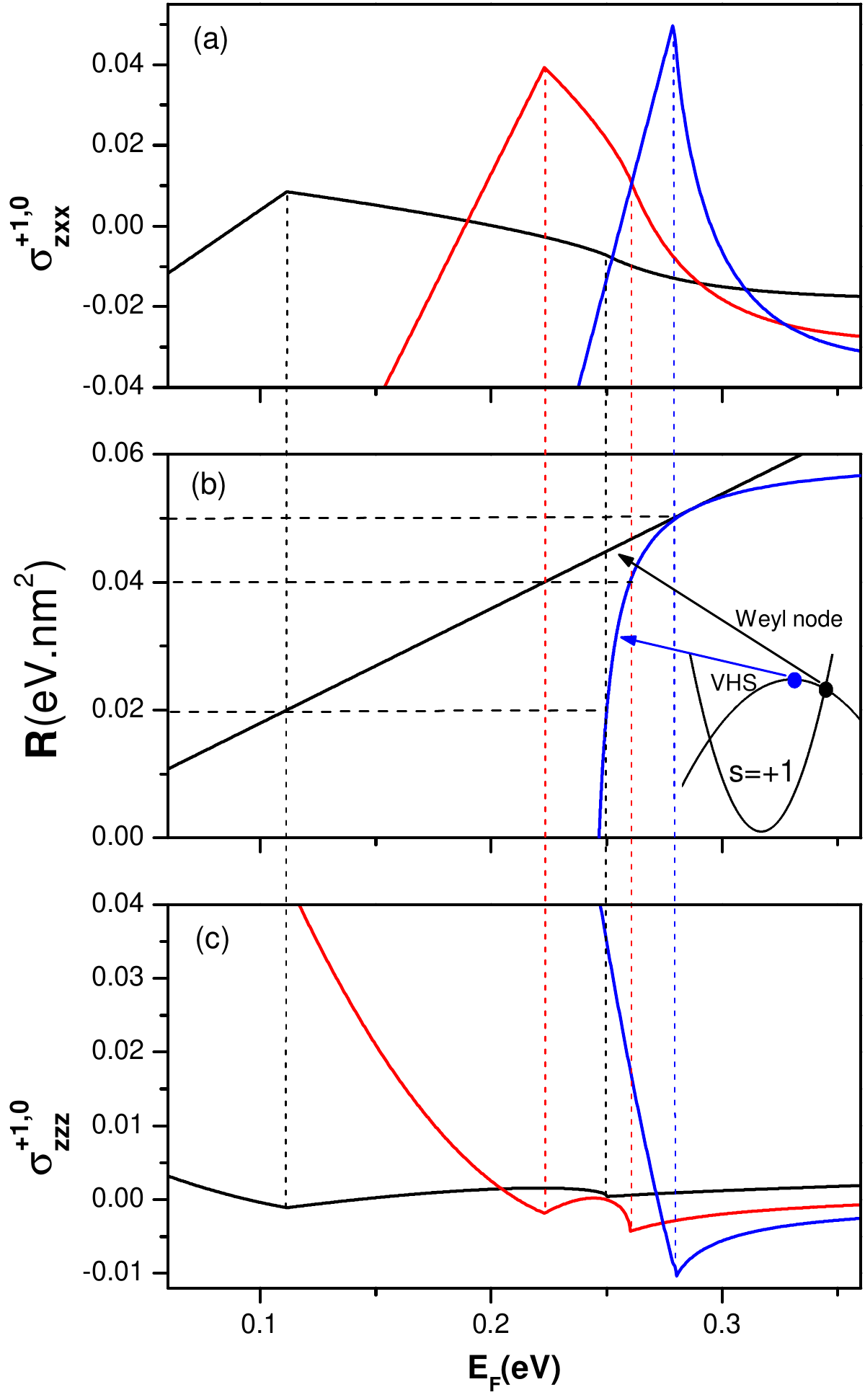}
\caption{(color online)
The second-order nonlinear conductivities $\sigma_{zxx}^{+1,0}$(a) and $\sigma_{zzz}^{+1,0}$(c) in unit of $e^2v_F\kappa _1/8\pi ^2\hbar ^2$ as a function of the Fermi energy for the tilts $R=0.02eV\cdot nm^2$(black line), $R=0.04eV\cdot nm^2$(red line), and $R=0.05eV\cdot nm^2$(blue line). Figure (b) depicts the tilt dependence of the Weyl node and VHS that are indicated by the energy band diagram in the inset of Figure(b). The specific positions of conductivity curve features, such as peaks and inflection points, in (a) and (c) are precisely identified by the intersecting horizontal and vertical dashed lines. The other parameters are taken as $V_I=0.04eV\cdot nm$, $Q_D=2nm^{-1}$, $J_{ex}=0eV$, and $\lambda=0.06nm$.}
\label{fig2}
\end{figure}

We begin our numerical analysis by examining the scenario where the parameter $J_{ex}$ is set to zero. In Fig. \ref{fig2}(a) and (c), the conductivity components $\sigma_{zxx}^{+1,0}$ and $\sigma_{zzz}^{+1,0}$ are presented across the VHS and the Weyl node (indicated by circles in the inset of Fig. \ref{fig2}(b)) within the $s=+1$ band, considering various tilt values represented by $R$.
Near the Weyl node, $\sigma_{zxx}^{+1,0}$ demonstrates a linear correlation with the Fermi energy $E_F$, resulting in a cusp peak concerning the Fermi energy at the Weyl node. This finding aligns with the outcomes of the single-node model\cite{Rui-Hao21,Gaoyang22}. Conversely, at the VHS, $\sigma_{zxx}^{+1,0}$ shows an inflection in relation to the Fermi energy, marking a transition from a linear to a nonlinear relationship. As $R$ increases, both the peak and the inflection shift towards higher energies. Fig. \ref{fig2}(b) precisely tracks their positions, depicting the evolution of the Weyl node and VHS influenced by the tilt. Remarkably, as the tilt strength amplifies, the Weyl node experiences a more rapid shift, causing the conductivity peak to approach the inflection point at the VHS (refer to Fig. \ref{fig2}(a)). Upon reaching a sufficient tilt magnitude, the Weyl node and VHS merge into a single point. In this scenario, the band structure around the Weyl node tends to flatten, leading to a rapid increase in the density of states and consequently enhancing electron nonlinear transport. This enhancement manifests in the increased conductivity peak in Fig. \ref{fig2}(a).

Moving to Fig. \ref{fig2}(c), we observe that the second-order nonlinear conductivity $\sigma_{zzz}^{+1,0}$ displays a dip either at the Weyl node or the VHS. Due to the distortion of the Fermi surface induced by the tilt, these dip structures exhibit asymmetry. With an increase in tilt, these two dips diminish in magnitude, and their separation reduces, resulting in the emergence of a broad peak between them. At the juncture where the Weyl node and VHS merge, this peak notably disappears, leading to a substantial dip.

\begin{figure}[htbp]
\centering
\includegraphics[width=7.6cm]{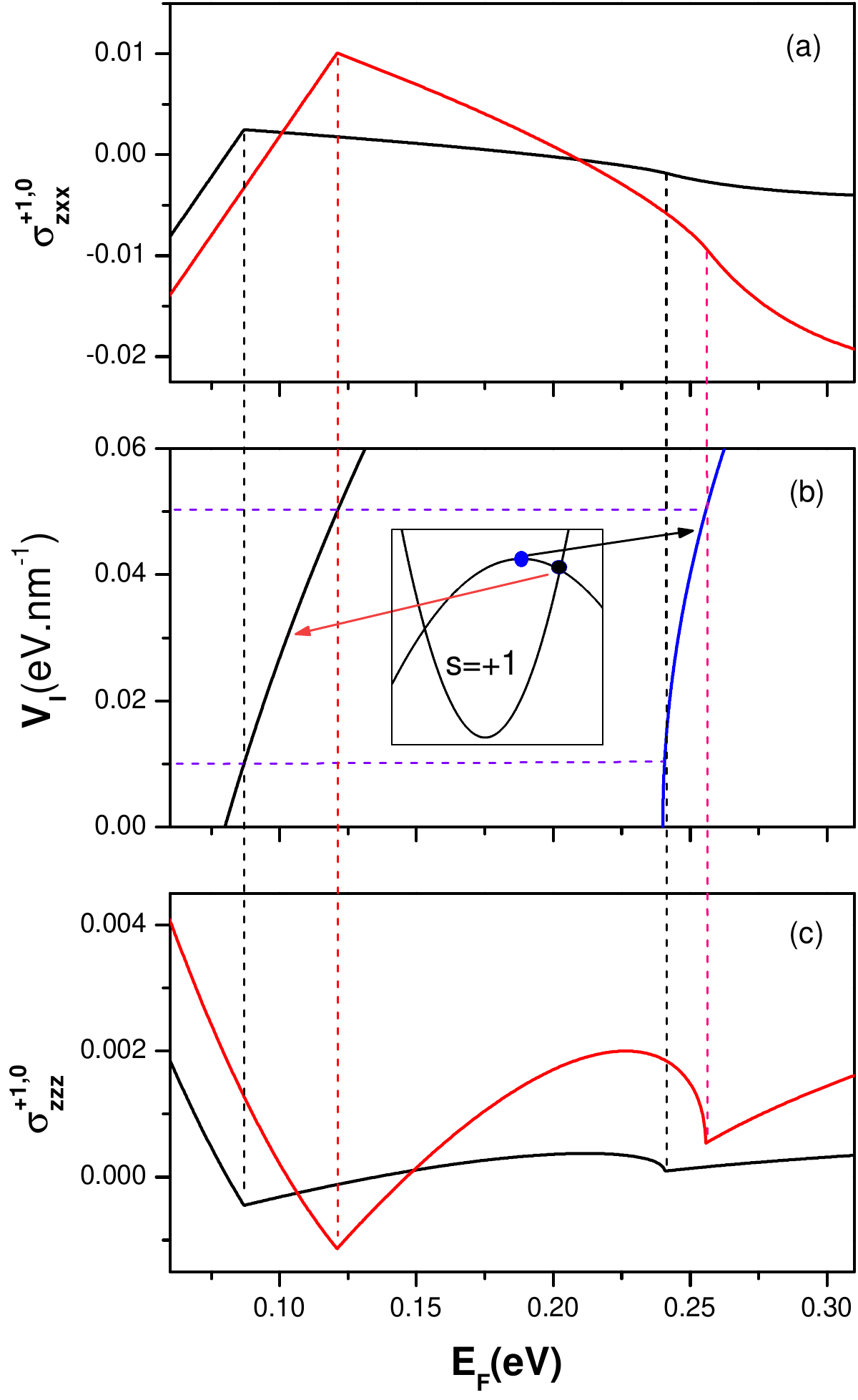}
\caption{(color online)
The second-order nonlinear conductivities $\sigma_{zxx}^{+1,0}$(a) and $\sigma_{zzz}^{+1,0}$(c) in unit of $e^2v_F\kappa _1/8\pi ^2\hbar ^2$ as a function of the Fermi energy for the parameter $V_I=0.01eV\cdot nm$(black line) and $V_I=0.05eV\cdot nm$(red line) at $R=0.02eV\cdot nm^{2}$.  Figure (b) shows the variation of the Weyl node and VHS, marked by the inset of Fig. (b), with the parameter $V_I$. The precise locations of salient features in the conductivity curves, including peaks and inflection points, in panels (a) and (c) are indicated by the intersection of horizontal and vertical dashed lines.
All other parameters remain consistent with those of Fig. \ref{fig2}.}
\label{fig3}
\end{figure}
By varying the parameter $V_I$, we analyze the second-order nonlinear conductivity components, $\sigma_{zxx}^{+1,0}$ and $\sigma_{zzz}^{+1,0}$ as a function of the Fermi energy $E_F$ in Fig. \ref{fig3}(a) and (c). Notably, the amplitudes of $\sigma_{zxx}^{+1,0}$ or $\sigma_{zzz}^{+1,0}$ near the Weyl node and VHS exhibit inconsistent changes with $V_I$. This behavior markedly differs from that observed when adjusting the tilt (as depicted in Fig. \ref{fig2}). This discrepancy stems from the distinct evolution of the Fermi surface induced by $R$ and $V_I$. Under variations in $V_I$, the Weyl node and VHS move at nearly identical rates, as illustrated in Fig. \ref{fig3}(b). Consequently, the separation between the Weyl node and the VHS remains constant, preserving the symmetric features of the Fermi surface. Conversely, when influenced by $R$, the symmetry of the Fermi surface is disrupted, indicated by the convergence of the Weyl node($W_{R,+1}$) and the VHS towards each other (refer to Fig.\ref{fig2}(b)).

\begin{figure}[htbp]
\centering
\includegraphics[width=7.6cm]{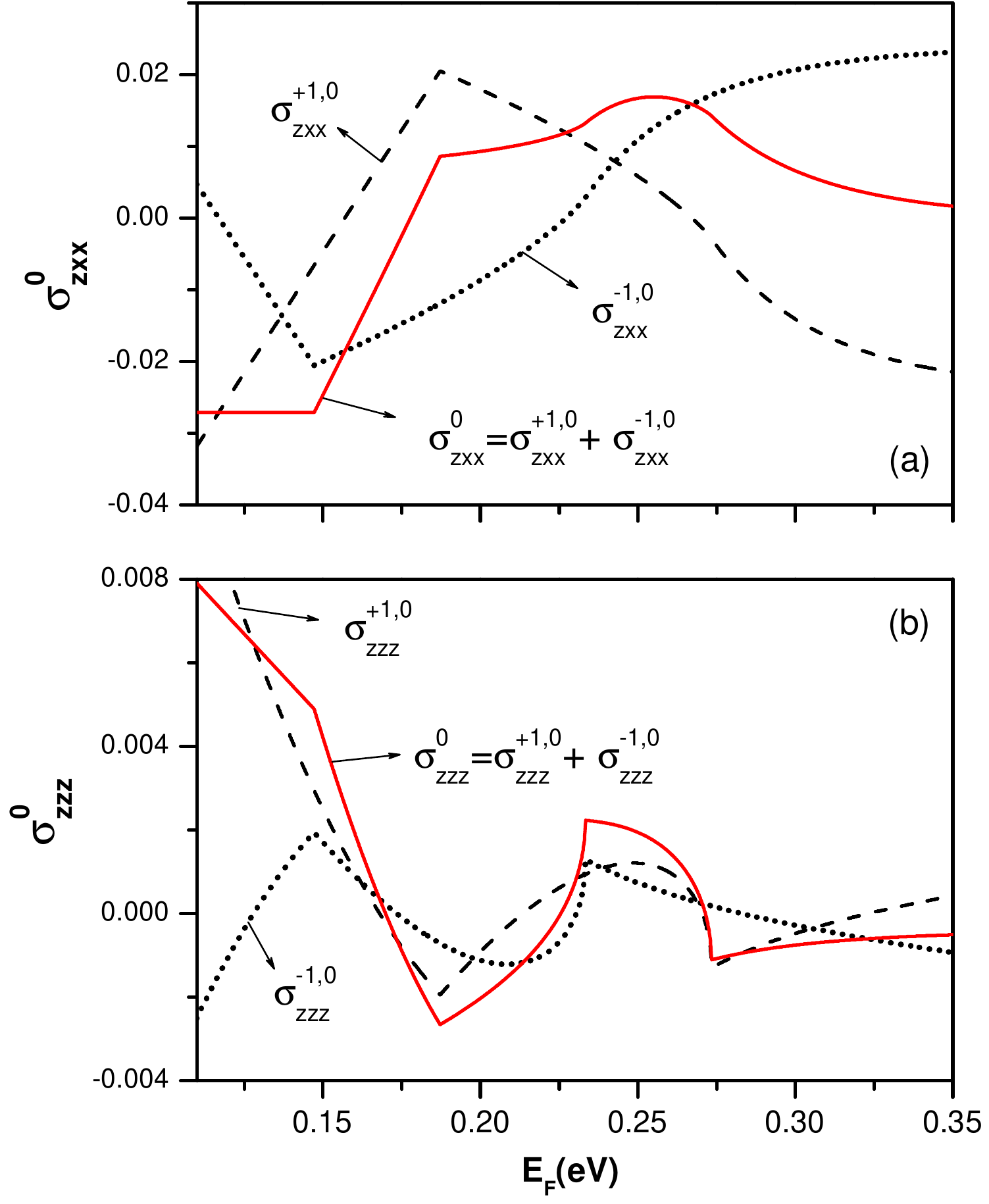}
\caption{(color online) The second-order nonlinear conductivities for the $s=+1$(dashed line) and $s=-1$(dotted line) bands, and their summation (solid line), measured in unit of $e^2v_F\kappa _1/8\pi ^2\hbar ^2$ as a function of the Fermi energy at $V_I=0.05 eV\cdot nm$, $R=0.02eV\cdot nm^{2}$ and $J_{ex}=0.02eV$.
 The other parameters are taken as the same as those of Fig.\ref{fig2}.}
\label{fig4}
\end{figure}

Indeed, at $J_{ex}=0$, due to opposite chiralities, the contributions of the $s=+1$ and $s=-1$ bands to the second-order conductivity tend to counteract each other. However, when ${J_{ex}} \neq 0$, the chirality symmetry breaks, and the $s=\pm1$ bands separate by $J_{ex}$. Consequently, the total second-order nonlinear conductivities display non-zero values and introduce additional inflection points compared to the case of a single $s=+1$ band (see Fig.\ref{fig4}). These intervals are determined by the relative displacements between the Weyl nodes and the VHS, showcasing the subtle relationship between band structure and electron transport.

\subsubsection{Second-order anomalous Hall conductivity $\sigma_{abc}^{s,H}$}

Based on the the parity of the velocity $\mathbf{v}_\alpha^s$ and Berry curvature $\mathbf{\Omega}_\alpha^s$, alongside the involvement of the antisymmetric tensor $\varepsilon_{acd}$,  the distinct component indices among $x$, $y$ and $z$ are requisite for the existence of $\sigma_{abc}^{s,H}$. By a straightforward calculation, the expressions are derived as follows:
\begin{footnotesize}
\begin{equation}
\begin{aligned}\label{hhd3}
&\sigma _{xyz}^{s,H} =  - \frac{se\kappa _2}{16\pi ^2\hbar}\int {d{k_z}} \frac{\Delta_s'(k_z)}{r (k_z)}[1- \frac{\Delta_s^2 (k_z)}{r^2 (k_z)}] ,
\end{aligned}
\end{equation}
\end{footnotesize}
\begin{footnotesize}
\begin{equation}
\begin{aligned}\label{hhd32}
&\sigma _{xzy}^{s,H} = {{{se\kappa _2}} \over {8\pi ^2\hbar}}\int dk_z\frac{\Delta_s (k_z)}{r^2 (k_z)} [{{{\Delta_s ({k_z})}\Delta_s'({k_z})} \over {{r ({k_z})}}} + {{2R{k_z}} \over {\hbar v_F}}].
\end{aligned}
\end{equation}
\end{footnotesize}
Additional nonzero components adhere to the relations: $\sigma _{zxy}^{s,H} =  - \sigma _{zyx}^{s,H} =  - \sigma _{yxz}^{s,H} = \sigma _{xyz}^{s,H}$ and $ - \sigma _{yzx}^{s,H} = \sigma _{xzy}^{s,H}$. The frequency dependence of conductivity, as expressed in Eqs. (\ref{hhd3}) and (\ref{hhd32}), is encapsulated within the factor $\kappa_2$. Under the transport limit, where $\omega\tau << 1$, the conductivity components $\sigma _{xyz}^{s,H}$ and $\sigma _{xzy}^{s,H}$ exhibit a direct proportionality to the relaxation time $\tau$ and remain invariant with respect to frequency. Conversely, under the optical or pristine limit, these conductivities, $\sigma _{xyz}^{s,H}$ and $\sigma _{xzy}^{s,H}$, become purely imaginary and demonstrate a proportionality to $1/\omega$, thereby reflecting the characteristics of second harmonic generation.

\begin{figure}[htbp]
\centering
\includegraphics[width=8.2cm]{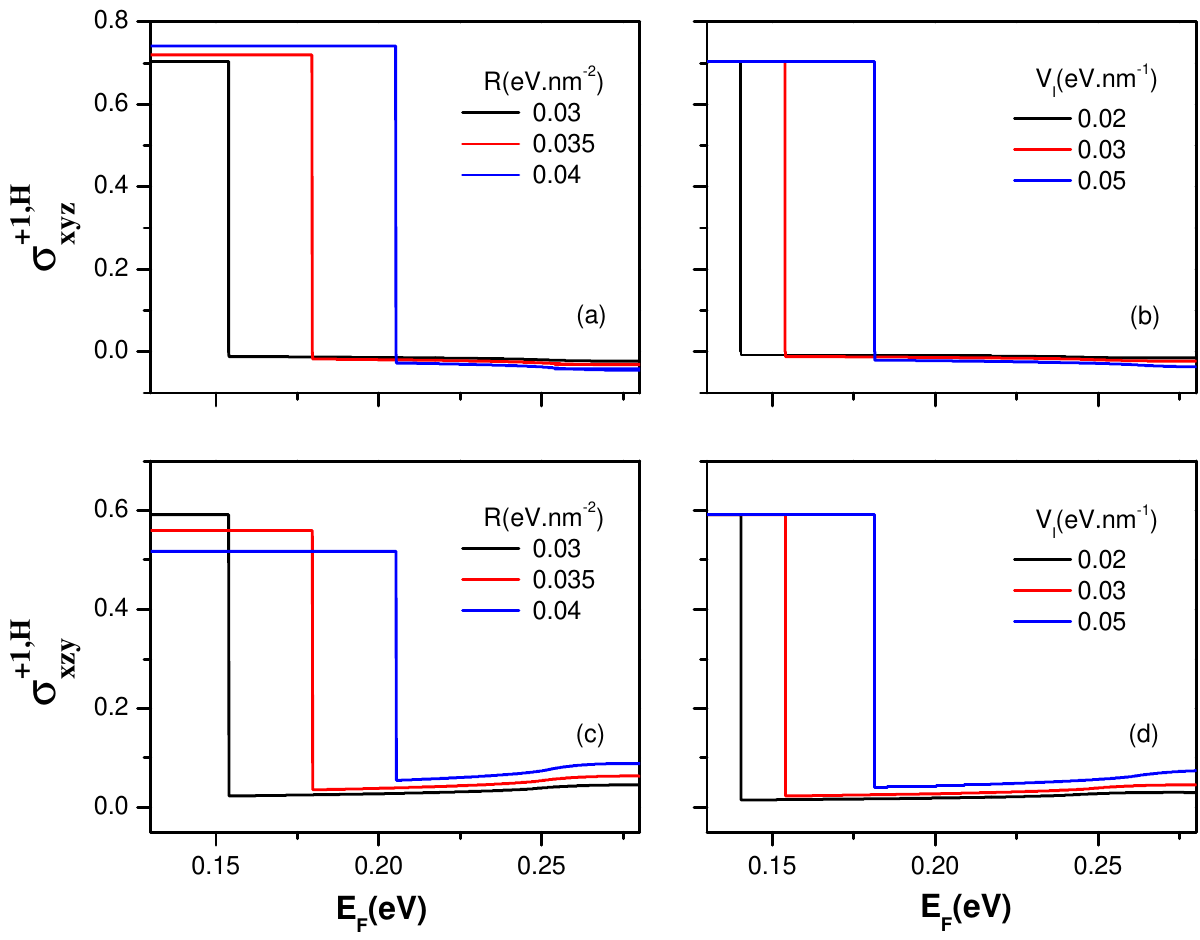}
\caption{(color online)
Depiction of second-order nonlinear conductivities ($\sigma_{xyz}^{+1,H}$ in (a) and (b), $\sigma_{xzy}^{+1,H}$ in (c) and (d)) for the $s=+1$ band, varying with $R$ and $V_I$. These conductivities, measured in units of $e\kappa_2/8\pi^2\hbar$, are presented as a function of the Fermi energy at $V_I=0.03eV\cdot nm$ (a) and (c), $R=0.03eV\cdot nm^2$ (b) and (d). Remaining parameters are consistent with those specified in Fig.\ref{fig2}.}
\label{fig5}
\end{figure}
\begin{figure}[htbp]
\centering
\includegraphics[width=7.2cm]{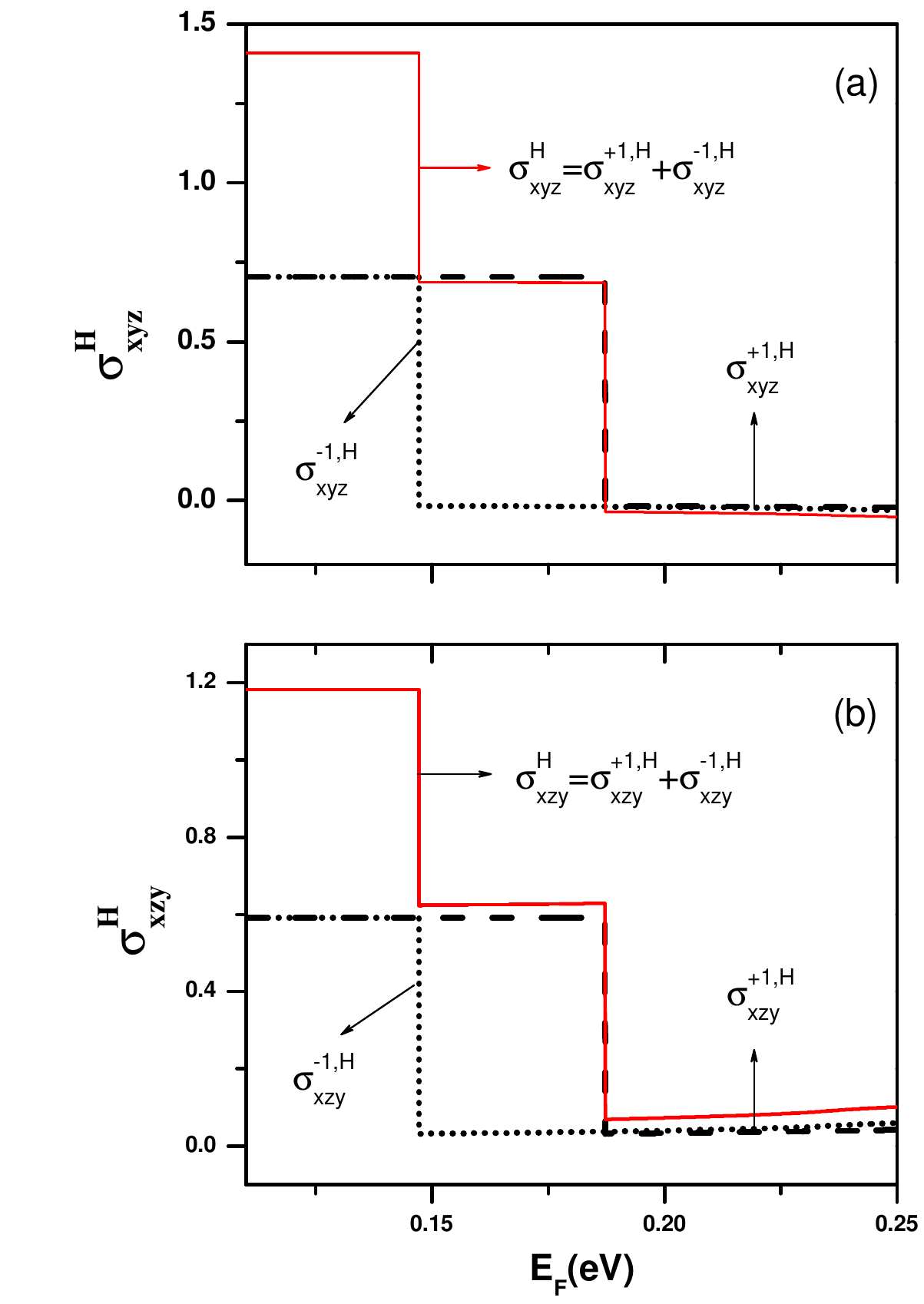}
\caption{(color online) The second-order nonlinear conductivities for  the $s=+1$(dashed line) and $s=-1$(dotted line) bands,
and their summation(solid line), measured in unit of $e^2v_F\kappa_1/8\pi^2\hbar^2$ as a function of the Fermi energy at $V_I=0.05eV\cdot nm$, $R=0.02eV\cdot nm^2$ and $J_{ex}=0.02eV$.
 The other parameters are taken as the same as those of Fig.\ref{fig2}.}
\label{fig6}
\end{figure}

Fig.\ref{fig5} depicts the behavior of the second-order anomalous Hall conductivities $\sigma _{xyz}^{+1,H}$ and $\sigma_{xzy}^{+1,H}$ as functions of the Fermi energy for different values of $R$ and $V_I$ at $J_{ex}=0$.
Both conductivities exhibit a plateau-like structure. For $\sigma_{xyz}^{+1,H}$, there's either an increase or decrease at low or high energies, respectively, with increasing $R$, while $\sigma_{xzy}^{+1,H}$ exhibits a inverse dependence on the $R$.  However, with an increase in $V_I$, $\sigma_{xyz}^{+1,H}$ ($\sigma_{xzy}^{+1,H}$) maintains a nearly constant plateau height at low energy, and slightly decrease(increases) at high energy. These distinct variations attributed to $R$ and $V_I$ are associated with the observed asymmetry in the Fermi surface (as depicted in Fig. \ref{Fig1}).
The plateau widths are determined by the separation between the Weyl nodes and VHS, hence their changes with increasing $R$ and $V_I$ display inconsistent behaviors. Upon the introduction of $J_{ex}$, the $s=\pm1$ bands split due to $J_{ex}$, consequently altering the plateau structure and resulting in the emergence of new plateaus in the total second-order nonlinear Hall conductivities, as illustrated in Fig. \ref{fig6}.

\subsection{Magnetic field induced contribution to second-order conductivity}

In the presence of a weak magnetic field, incorporating the effect of the orbital magnetic moment modifies the distribution function, expressed as\cite{Shudan2016,Gaoyang22}
\begin{equation}
{f_0} = f_0^s - {{\bf{m}}^s_{\alpha}} \cdot {\bf{B}}{{\partial f_0^s} \over {\partial \varepsilon _{\alpha}^s}}.
\label{f0}
\end{equation}
By substituting Eq.\eqref{mkds}, Eq.\eqref{f1} and Eq.\eqref{f0} into the Eq.\eqref{cdestit}, we only keep the linear term in ${\bf{B}}$:
\begin{footnotesize}
\begin{equation}
\begin{array}{lll}
\mathbf{j}^{(B)s} &=&  - {\kappa _1}\int {[d{\bf{k}}]} [{\bf{v}}_{\alpha}^s - {\nabla _{\bf{k}}}({\bf{m}}_{\alpha}^s \cdot {\bf{B}}) +e({\bf{v}}_{\alpha}^s \cdot {\bf{\Omega }}_{\alpha}^s){\bf{B}}]C_{\mathbf{k}}\mathbf{\Lambda}_{\mathbf{k}}\\
&& \cdot \nabla_{\mathbf{k}}[ C_{\mathbf{k}}\mathbf{\Lambda}_{\mathbf{k}}
 \cdot\nabla_{\mathbf{k}}(f_0^s - {\bf{m}}_{\alpha}^s \cdot {\bf{B}}{{\partial f_0^s} \over {\partial \varepsilon _{\alpha}^s}})] \\
&& - \kappa _2\int {[d{\bf{k}}]} {\bf{E}} \times {\bf{\Omega }}_{\alpha}^sC_{\mathbf{k}}\mathbf{\Lambda}_{\mathbf{k}}
 \cdot \nabla_{\bf{k}}(f_0^s - {\bf{m}}_{\alpha}^s \cdot {\bf{B}}{{\partial f_0^s} \over {\partial \varepsilon _{\alpha}^s}}),
\end{array}
\label{cde}
\end{equation}
\end{footnotesize}
where $\Lambda_{\mathbf{k}}=e{\bf{E}} + {{{e^2}} \over \hbar }({\bf{E}} \cdot {\bf{B}}){\bf{\Omega }}_{\alpha}^s$, and $C_{\mathbf{k}}=1 - {e \over \hbar }({\bf{B}} \cdot {\bf{\Omega }}_{\alpha}^s)$. Obviously, in Eq.\eqref{cde}, the direction of the current density contributed by the first integral term is governed by a combination of the wave packet velocity ${\bf{v}}_{\alpha}^s$, the Berry curvature ${\bf{\Omega }}_{\alpha}^s$, and the orbital magnetic moment  ${\bf{m}}_{\alpha}^s $. While  the direction of the current density contributed by the second integral term is solely determined by the coupling term of the electric field  and Berry curvature ${\bf{E}}\times{\bf{\Omega }}_{\alpha}^s$, resulting in a nonlinear Hall current. We also express Eq.\eqref{cde} in the form of $j_a^{(B)s}=\sigma_{abc}^{(B)s}E_bE_c$, where $\sigma_{abc}^{(B)s}$ represents the magnetic field induced contribution to the second-order nonlinear conductivity.

When the applied magnetic field ${\bf{B}}$ aligns with the electric field ${\bf{E}}$ (i.e., ${\bf{E}} \cdot {\bf{B}} \ne 0$), evidently, the second integral term in Eq.\eqref{cde} does not contribute to the second-order conductivity. Substituting Eqs.\eqref{berrv} and \eqref{mk} into Eq.\eqref{cde}, we derive the following expressions from the first integral terms of Eq.\eqref{cde}:
\begin{footnotesize}
\begin{equation}
\begin{aligned}\label{xxx}
&\sigma _{xxx}^{(B)s} = {\sigma _1}{B_x}\int {[d{\bf{k}}} ]\{ {{\partial {v}_{\alpha x}^{s2}} \over {2\partial {k_x}}}\Omega _{\alpha x}^s + [{{\partial {v}_{\alpha x}^{s2}} \over {2\partial {k_y}}} + {{\partial (v_{\alpha x}^sv_{\alpha y}^s)} \over {\partial {k_x}}}]\Omega _{\alpha y}^s\\
& + [{{\partial {v}_{\alpha x}^{s2}} \over {2\partial {k_z}}} + {{\partial (v_{\alpha x}^sv_{\alpha z}^s)} \over {\partial {k_x}}}]\Omega _{\alpha z}^s + {v_{\alpha x}^{s2}}\nabla\cdot\mathbf{\Omega}_\alpha^s+ v_{\alpha x}^sv_{\alpha y}^s{{\partial \Omega _{\alpha y}^s} \over {\partial {k_x}}} \\
&  + v_{\alpha x}^sv_{\alpha z}^s{{\partial \Omega _{\alpha z}^s} \over {\partial {k_x}}} - {{{\partial ^2}m_{\alpha x}^s} \over {e\partial k_x^2}}v_{\alpha x}^s + {{{\partial ^2}v_{\alpha x}^s} \over {e\partial k_x^2}}m_{\alpha x}^s\} {{\partial f_0^s} \over {\partial {\varepsilon _{\alpha}^s}}},
\end{aligned}
\end{equation}
\end{footnotesize}
\begin{footnotesize}
\begin{equation}
\begin{aligned}\label{zzz}
& \sigma _{zzz}^{(B)s} = {\sigma _1}{B_z}\int {[d{\bf{k}}} ]\{ {{\partial {v}_{\alpha z}^{s2}} \over {2\partial {k_z}}}\Omega _{\alpha z}^s + [{{\partial {v}_{\alpha z}^{s2}} \over {2\partial {k_y}}} + {{\partial (v_{\alpha z}^sv_{\alpha y}^s)} \over {\partial {k_z}}}]\Omega _{\alpha y}^s \\
& + [{{\partial {v}_{\alpha z}^{s2}} \over {2\partial {k_x}}} + {{\partial (v_{\alpha x}^sv_{\alpha z}^s)} \over {\partial {k_z}}}]\Omega _{\alpha x}^s + v_{z\alpha }^{s2}\nabla\cdot\mathbf{\Omega}_\alpha^s+ v_{x\alpha }^sv_{z\alpha }^s{{\partial \Omega _{x\alpha }^s} \over {\partial {k_z}}}\\
& + v_{y\alpha }^sv_{z\alpha }^s{{\partial \Omega _{y\alpha }^s} \over {\partial {k_z}}} - {{{\partial ^2}m_{\alpha z}^s} \over {e\partial k_z^2}}v_{\alpha z}^s + {{{\partial ^2}v_{\alpha z}^s} \over {e\partial k_z^2}}m_{\alpha z}^s\} {{\partial f_0^s} \over {\partial {\varepsilon _{\alpha}^s}}},
\end{aligned}
\end{equation}
\end{footnotesize}
where ${\sigma _1} = {{e^4\tau ^2v_F^2} \over {(1 - 2i\omega \tau )(1 - i\omega \tau ){h ^2}}}$. Notably, due to the symmetry of the $k_x$-axis and the $k_y$-axis, $\sigma _{yyy}^{(B)s}({B_y})/B_y = \sigma _{xxx}^{(B)s}(B_x)/B_x$. All other components are rendered as zero.

By orienting the magnetic field ${\bf{B}}$ perpendicular to the electric field ${\bf{E}}$ (i.e., ${\bf{E}} \cdot {\bf{B}} = 0$), we ascertain the nonlinear Hall conductivity components in the following manner:
\begin{footnotesize}
\begin{equation}\label{xyy}
\begin{aligned}
&\sigma _{xyy}^{(B)s} = {\sigma _1}{B_x}\int {[d{\bf{k}}} ][ {{\partial v_{\alpha x }^s} \over {\partial {k_y}}}v_{\alpha y }^s\Omega _{\alpha x }^s - {{\partial (v_{\alpha y }^s\Omega _{\alpha y }^s + v_{\alpha z }^s\Omega _{\alpha z }^s)} \over {\partial {k_y}}}v_{\alpha y }^s\\
&  - {{\partial^2 m_{\alpha x }^s} \over {e\partial {k_x}\partial {k_y}}}v_{\alpha y }^s + {{{\partial ^2}v_{\alpha x }^s} \over {e\partial {k_y^2}}}m_{\alpha x }^s]{{\partial f_0^s} \over {\partial {\varepsilon _\alpha^{s}}}},
\end{aligned}
\end{equation}
\end{footnotesize}
\begin{footnotesize}
\begin{equation}\label{xzz}
\begin{aligned}
&\sigma _{xzz}^{(B)s} = {\sigma _1}{B_x}\int {[d{\bf{k}}]} [ -{{\partial v_{\alpha x }^s} \over {\partial {k_z}}} v_{\alpha z }^s\Omega _{\alpha x }^s + {{\partial (v_{\alpha y }^s\Omega _{\alpha y }^s + v_{\alpha z }^s\Omega _{\alpha z }^s)} \over {\partial {k_z}}}v_z^s\\
& - {{\partial m_{\alpha x }^s} \over {\partial {k_x}\partial {k_z}}}v_{z\alpha }^s + {{{\partial ^2}v_{\alpha x }^s} \over {\partial {k_z^2}}}m_{\alpha x }^s]{{\partial f_0^s} \over {\partial {\varepsilon _{\alpha}^s}}},
\end{aligned}
\end{equation}
\end{footnotesize}
\begin{footnotesize}
\begin{equation}\label{zyy}
\begin{aligned}
&\sigma _{zyy}^{(B)s} = \sigma _1B_z\int [d\mathbf{k}] [ -{{\partial v_{\alpha z }^s} \over {\partial k_y}}v_{\alpha y }^s\Omega _{\alpha z }^s + {{\partial (v_{\alpha x }^s\Omega _{\alpha x }^s + v_{\alpha y }^s\Omega _{\alpha y }^s)} \over {\partial {k_y}}}v_{\alpha y }^s\\
&  - {{{\partial ^2}m_{\alpha z }^s} \over {e\partial {k_z}\partial {k_y}}}v_{\alpha y }^s + {{{\partial ^2}v_{\alpha z }^s} \over {e\partial {k_y^2}}}m_{\alpha z }^s]{{\partial f_0^s} \over {\partial {\varepsilon _{\alpha}^s}}}.
\end{aligned}
\end{equation}
\end{footnotesize}
The additional non-zero conductivity components conform to relationships: $\sigma _{yxx}^{(B)s}/{B_y} = \sigma _{xyy}^{(B)s}/{B_x}=-\sigma _{zyy}^{(B)s}/{B_z}$, $\sigma _{zxx}^{(B)s} = \sigma _{zyy}^{(B)s}$, and $\sigma _{yzz}^{(B)s}/{B_y} = \sigma _{xzz}^{(B)s}/{B_x}$. Notably, Eqs.\eqref{xyy}, \eqref{xzz} and \eqref{zyy} stem from the contributions of the primary integral terms in Eq.\eqref{cde}.
The contribution arising from the second integral term in Eq.\eqref{cde} necessitates distinct component indices ($a, b, c$) for the conductivity components $\sigma _{abc}^{(B)s}$. Through further computation involving Eq.\eqref{cde}, we find
\begin{footnotesize}
\begin{equation}\label{Hzxy}
\begin{aligned}
&\sigma _{zxy}^{(B)s} =- {\sigma _2}{B_z}\int {[d{\bf{k}}]} ( {{\partial \Omega _{\alpha x }^s} \over {\partial {k_x}}}m_{\alpha z }^s + ev_{\alpha x }^s\Omega _{\alpha x }^s\Omega _{\alpha z }^s) {{\partial {f_0^s}} \over {\partial {\varepsilon _{\alpha}^s}}},
\end{aligned}
\end{equation}
\end{footnotesize}
where we define ${\sigma _2} = {{{e^4}\tau } \over {4\pi^2{\hbar ^3}(1 - i\omega \tau )}}$. The other non-zero conductivity components can be related as follows: $\sigma _{zxy}^{(B)s} =  - \sigma _{zyx}^{(B)s}$, $\sigma _{yzx}^{(B)s}/B_y =  - \sigma _{xzy}^{(B)s}/B_x=2\sigma _{zxy}^{(B)s}/B_z$. We

\begin{figure}[htbp]
\centering
\includegraphics[width=8.6cm]{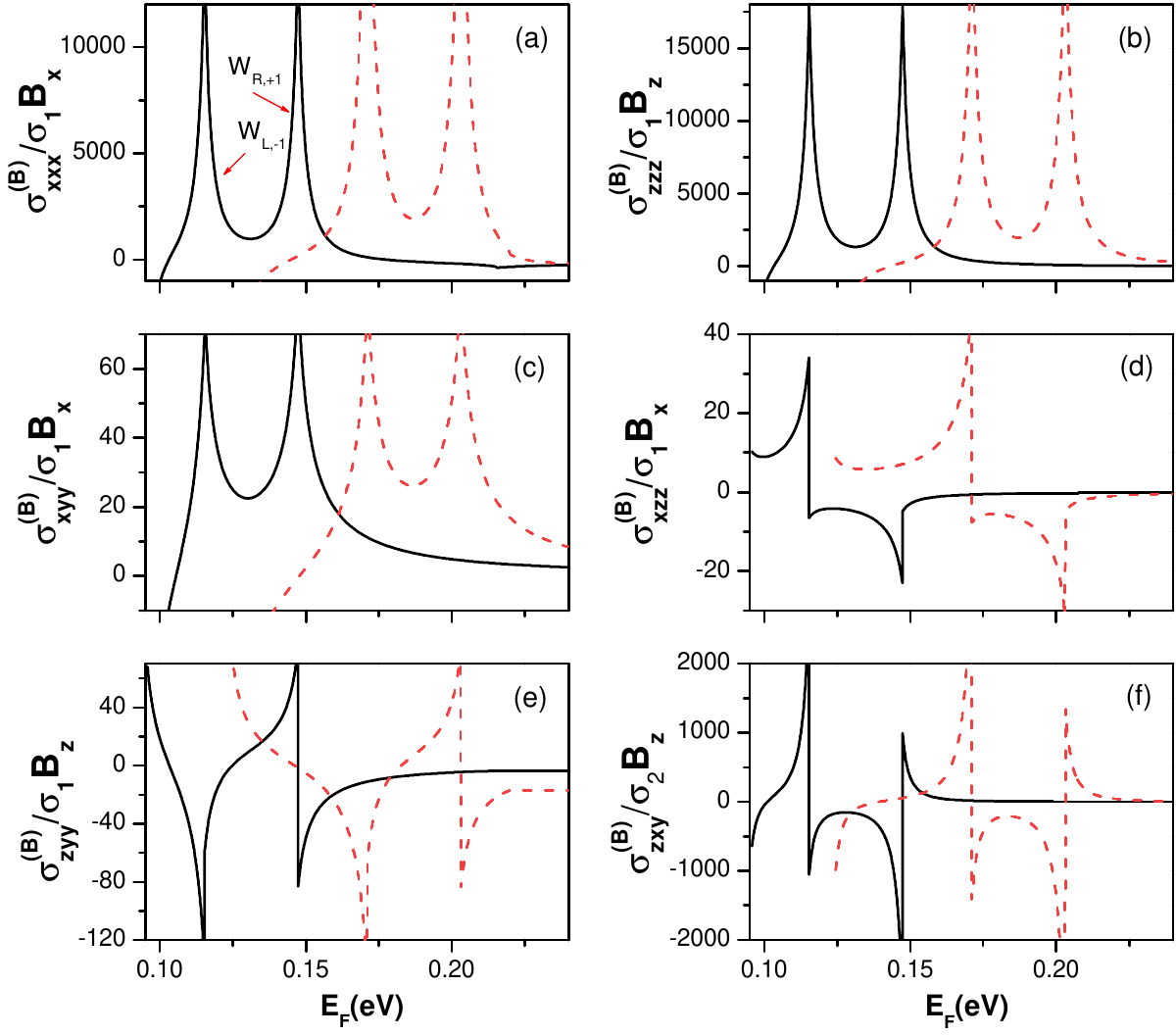}
\caption{(color online)
The total second-order nonlinear conductivity components of the $s=\pm1$ bands, namely $\sigma _{xxx}^{(B)}$ (a), $\sigma _{zzz}^{(B)}$ (b), $\sigma _{xyy}^{(B)}$ (c), $\sigma _{xzz}^{(B)}$ (d), $\sigma _{zyy}^{(B)}$ (e), and $\sigma _{zxy}^{(B)}$ (f), as a function of Fermi energy for two distinct values of $R=0.025eV\cdot nm^{2}$ (solid line) and $R=0.035eV\cdot nm^{2}$ (dashed line) at $J_{ex}=0.02eV$ and $V_I=0.04eV\cdot nm$. All other parameters remain consistent with those defined in Fig.\ref{fig2}.
}
\label{fig7}
\end{figure}
In the presence of tilt $R$ and the splitting term $J_{ex}$, the energies of the four Weyl nodes exhibit discrepancies, as illustrated in Fig. \ref{Fig1}. Assuming a comparative scale between the splitting energy $J_{ex}$ and the VHS energy, the ordering of Weyl node energies remains consistent: $E_{W_{R,+1}}>E_{W_{L,-1}}>E_{W_{L,+1}}>E_{W_{R,-1}}$.
Fig.\ref{fig7} illustrates the dependence of the total second-order nonlinear conductivity of the $s=\pm 1$ bands, $\sigma_{abc}^{(B)}=\sigma_{abc}^{(B)+1}+\sigma_{abc}^{(B)-1}$, induced by the magnetic field on the Fermi energy $E_F$ for varying tilts. The second-order conductivity components $\sigma_{aaa}^{(B)}(a=x,z)$ and $\sigma_{xyy}^{(B)}$ across the Weyl nodes $W_{R,+1}$ and $W_{L,-1}$ display distinct sharp peaks. The asymmetry in these peak structures concerning the Weyl nodes indicates the influence of the magnetic moment, which suppresses nonlinear electron transport while enhancing nonlinear hole transport, as detailed in Eqs. \eqref{xxx}-\eqref{xzz}. Terms related to the magnetic moment demonstrate positivity, while others rely on the band index $\alpha(=\pm1)$.
For the conductivity components $\sigma_{xzz}^{(B)}$, $\sigma_{zyy}^{(B)}$, and $\sigma_{zxy}^{(B)}$, the dominance of the magnetic moment becomes evident, resulting in a kinked structure showing a sudden shift from positive to negative values near the Weyl nodes $W_{R,+1}$ and $W_{L,-1}$, as depicted in Fig. \ref{fig7}(d),(e), and (f).
Further examination of the second-order conductivity expressions (Eqs. \eqref{xxx}-\eqref{Hzxy}) reveals that the peak structures follow an asymptotic relationship $\sigma_{aaa}^{(B)}, \sigma_{xyy}^{(B)} \propto \frac{1}{(E_F - E_{W_{\alpha,s}})^2}$ near the Weyl nodes $W_{\alpha,s}$. Meanwhile, the kinked structure (as shown in Fig. \ref{fig7}(d),(e), and (f)) adheres to the relation $\sigma_{xzz}^{(B)}, \sigma_{zyy}^{(B)}, \sigma_{zxy}^{(B)} \propto \pm\frac{\text{sgn}(E_F - E_{W_{\alpha,s}})}{(E_F - E_{W_{\alpha,s}})^2}$ near the Weyl nodes $W_{\alpha,s}$.
As the tilt $R$ increases, these characteristic structures shift towards higher energies while retaining their essential peak and kinked features. However, near the VHS, the distinct feature induced by the VHS is not clearly observed due to the suppressing effect of the magnetic moment, overriding the influence of the VHS. Even an enhancement in the VHS energy nearly fails to amplify its influence.
In analyzing the frequency dependence of conductivity, we observe a similarity between the magnetic field-induced contribution to the second-order conductivity and the case of zero magnetic field. Consequently, in the transport regime where $\omega\tau\ll 1$, both $\sigma_{abb}$ and $\sigma_{aaa}$ vary proportionally to $\tau^2$, whereas $\sigma_{abc}$ scales linearly with $\tau$. Conversely, in the optical limit where $\omega\tau\gg 1$, $\sigma_{abb}$ and $\sigma_{aaa}$ scale inversely with the square of frequency ($1/\omega^2$), and $\sigma_{aaa}$ possesses an imaginary component, obeying the relation $\sigma_{aaa}\propto 1/\omega$.

Moreover, from the aforementioned results, we can further assess the scale of the nonlinear magneto-optical susceptibility using the equation $\chi^{(2\omega)}=j/(i\omega)\epsilon_0E^2$\cite{Morimoto16}, where $\epsilon_0$ represents vacuum permittivity. By adopting the parameters $E_F=8 meV$, $v_F= 3.2 \times {10^5} m/s$, $V_I = 0.015 eV \cdot nm^{- 1}$, $J_{ex}=0.01 eV$, $\lambda  = 0.5 \text{nm}$, $R = 0.3  \text{eV} \cdot nm^2$, and $Q_D = 0.8 nm^{-1}$, we derive $\chi^{(2\omega )} \approx 5.6 \times {10^3}B$  $pm/V$ at $\omega = 8\pi THz$. Recent observations in WSMs have demonstrated significantly enhanced nonlinear optical responses, encompassing photocurrent\cite{manp2017}, second- or third-harmonic generation\cite{wunp2016}, and the optical Kerr effect\cite{arxivchoi}. Specifically, an extraordinarily high coefficient for the linear magneto-optic Kerr effect has been reported within a magnetic WSM\cite{higonp2018}. These observations indicate the potential realization of our findings concerning nonlinear magneto-optical features induced by the magnetic moment within magnetic WSMs, particularly within the infrared regime.

\section{Conclusions}
\label{sec:conclusions}
We study the nonlinear magneto-optical transport properties within non-centrosymmetric magnetic WSMs. We construct an effective low-energy model that incorporates essential elements: the VHS, a tilted term, and the presence of broken space inversion and broken time inversion terms.
Using this model, we derive analytical expressions for the second-order nonlinear conductivity components through the semiclassical Boltzmann equation. We observe that in the absence of a magnetic field, the second-order nonlinear Drude conductivity components 
display inflection or dip behaviors across the VHS. Conversely, the second-order nonlinear anomalous Hall conductivity, primarily influenced by the Berry curvature dipole, showcases a subtle plateau-like structure.
Significantly, intensifying the tilt strength amplifies these second-order conductivity features at the singularity due to the increased VHS energy.
Additionally, our investigation explores the magnetic field-induced impact on the second-order nonlinear conductivity, revealing that the resulting magnetic moment-induced suppression and enhancement of nonlinear electron and hole transport help counteract the influence of the VHS. This effect specifically generates an asymmetric peak or a kinked-like structure near the Weyl nodes.
Upon evaluating the magnitude of these second-order conductivity components, our findings suggest the potential observability of these phenomena in realistic magnetic WSMs.

\section*{ACKNOWLEDGMENTS}
We thank Dr. Yang Gao for the valuable discussions. This work was supported by the Postgraduate Scientific Research Innovation Project of Hunan Province (Grant No.CX20220958), the Natural
Science Foundation of Hunan Province, China (Grant No.
2023JJ30005), and the Open Research Fund of the Hunan Province Higher Education Key Laboratory of Modeling and Monitoring on the Near-Earth Electromagnetic Environments (GrantNo. N201904), Changsha University of Science and Technology.

\appendix
\section{ Density of states}\label{a}
The density of states for the $s=+1$ band can be calculated by
\begin{footnotesize}
\begin{equation}
\begin{aligned}
&\rho_{+1} (\varepsilon ) = \frac{1}{V}\sum\limits_{\mathbf{k}\alpha}\delta [\varepsilon  - \varepsilon_\alpha^{+1} (\mathbf{k})]
\end{aligned}\label{dos}
\end{equation}
\end{footnotesize}
To solve Eq.\eqref{dos}, we need to convert the summation over $\mathbf{k}$ into an integral in the three dimensional momentum space. After a straightforward calculation, we get the analytical expression for DOS as follows:

For $\varepsilon  > \varepsilon _{VHS}^c$,
\begin{footnotesize}
\begin{equation}
\begin{aligned}
&\rho _{+1}(\varepsilon ) = {1 \over {4{\pi ^2}v_F^2}}[(\varepsilon  - {J_{ex}})({k_{z3}} - {k_{z1}}) - {R \over 3}{({k_{z3}} - {k_{z1}})^3}].
\end{aligned}
\end{equation}
\end{footnotesize}
For $\varepsilon_{W_{R, + 1}} < \varepsilon  < \varepsilon _{VHS}^{\rm{c}}$,
\begin{footnotesize}
\begin{equation}
\begin{aligned}
&\rho _{+1}(\varepsilon ) = {1 \over {4{\pi ^2}v_F^2}}\{ (\varepsilon  - {J_{ex}})({k_{z2}} + {k_{z3}} - {k_{z1}} - {k_{z4}})\\
& - {R \over 3}[{({k_{z2}} - {k_{z1}})^3} + {({k_{z3}} - {k_{z4}})^3}]\}  .
\end{aligned}
\end{equation}
\end{footnotesize}
For $\varepsilon_{W_{L, + 1}} < \varepsilon  < \varepsilon_{W_{R, + 1}}$,
\begin{footnotesize}
\begin{equation}
\begin{aligned}
&{\rho _{+1}}(\varepsilon ) = {1 \over {4{\pi ^2}v_F^2}}\{ (\varepsilon  - {J_{ex}})({k_{z2}} + {k_{z4}} - {k_{z1}} - {k_{z3}})\\
&- {R \over 3}[{({k_{z2}} - {k_{z1}})^3} + {({k_{z4}} - {k_{z3}})^3}]\} .
\end{aligned}
\end{equation}
\end{footnotesize}
For $\varepsilon _{VHS}^{\rm{v}} < \varepsilon  < \varepsilon_{W_{{\rm{L}}, + 1}}$,
\begin{footnotesize}
\begin{equation}
\begin{aligned}
&\rho _{+1}(\varepsilon ) = -{1 \over {4{\pi ^2}v_F^2}}\{ (\varepsilon  - {J_{ex}})({k_{z1}} + {k_{z4}} - {k_{z2}} - {k_{z3}})\\
&  - {R \over 3}[{({k_{z1}} - {k_{z2}})^3} + {({k_{z4}} - {k_{z3}})^3}]\}.
\end{aligned}
\end{equation}
\end{footnotesize}
For $\varepsilon  < \varepsilon _{VHS}^v$,
\begin{footnotesize}
\begin{equation}
\begin{aligned}
&\rho _{+1}(\varepsilon ) = -{1 \over {4{\pi ^2}v_F^2}}[ (\varepsilon  - {J_{ex}})({k_{z4}} - {k_{z2}}) - {R \over 3}{({k_{z4}} - {k_{z2}})^3}].
\end{aligned}
\end{equation}
\end{footnotesize}
where
\begin{footnotesize}
\begin{equation}\label{A5}
\begin{aligned}
&{k_{z1}} = {{{v_F}\lambda q - \sqrt {{{({v_F}\lambda q)}^2} + ({v_F}\lambda  + R)(\varepsilon  - {J_{ex}} + {v_F}\lambda Q_D^2)} } \over {R + {v_F}\lambda }},\\
\end{aligned}
\end{equation}
\end{footnotesize}
\begin{footnotesize}
\begin{equation}\label{A7}
\begin{aligned}
&{k_{z2}} = {{ - {v_F}\lambda q + \sqrt {{{({v_F}\lambda q)}^2} + (R - {v_F}\lambda )(\varepsilon  - {J_{ex}} - {v_F}\lambda Q_D^2)} } \over {R - {v_F}\lambda }},\\
\end{aligned}
\end{equation}
\end{footnotesize}
\begin{footnotesize}
\begin{equation}\label{A5}
\begin{aligned}
&{k_{z3}} = {{{v_F}\lambda q + \sqrt {{{({v_F}\lambda q)}^2} + ({v_F}\lambda  + R)(\varepsilon  - {J_{ex}} + {v_F}\lambda Q_D^2)} } \over {R + {v_F}\lambda }},
\end{aligned}
\end{equation}
\end{footnotesize}
\begin{footnotesize}
\begin{equation}\label{A7}
\begin{aligned}
&{k_{z4}} = {{ - {v_F}\lambda q - \sqrt {{{({v_F}\lambda q)}^2} + (R - {v_F}\lambda )(\varepsilon  - {J_{ex}} - {v_F}\lambda Q_D^2)} } \over {R - {v_F}\lambda }}.
\end{aligned}
\end{equation}
\end{footnotesize}
Additionally, the VHS energies for the conduction band and valence band are expressed as $\varepsilon _{VHS}^c = {{{{({v_F}\lambda q)}^2}} \over {{v_F}\lambda  - R}} + {v_F}\lambda Q_D^2 + {J_{ex}}$, and $\varepsilon _{VHS}^v =  - {{{{({v_F}\lambda q)}^2}} \over {{v_F}\lambda  + R}} - {v_F}\lambda Q_D^2 + {J_{ex}}$. In the presence of the tilt, the energies of the Weyl nodes are modified to
$\varepsilon_{{\rm{W}}_{R, + 1}} = {(\sqrt {Q_D^2 + {q^2}}  + q)^2} + {J_{ex}}$, and $\varepsilon_{{\rm{W}}_{L, + 1}} = {(\sqrt {Q_D^2 + {q^2}}  - q)^2} + {J_{ex}}$.

\end{document}